\documentclass[12pt]{article}
\usepackage{graphicx}
\usepackage{eepic}
\usepackage{epic}

\newcommand{\beq}{\begin{equation}}
\newcommand{\eeq}{\end{equation}}
\def\bea{\begin{eqnarray}}
\def\eea{\end{eqnarray}}
\def\nn{\nonumber}

\begin{document}

\newcommand{\sheptitle}{ A Possibility of large electro-weak penguin 
                         contribution \\ in $B\rightarrow K\pi$ modes }
\newcommand{\shepauthor}{Tadashi Yoshikawa}
\newcommand{\shepaddressKEK}{
            Theory Group, KEK
                       Tsukuba, 305-0801, Japan.}

\date{\today}

\begin{titlepage}
\begin{flushright}
hep-ph/0306147\\
KEK-TH-894\\
\today
\end{flushright}
\vspace{.1in}
\begin{center}
{\large{\bf \sheptitle}}
\bigskip \medskip \\
\shepauthor \\
\mbox{} \\
\medskip
{\it \shepaddressKEK} \\
\vspace{.5in}

\bigskip \end{center} \setcounter{page}{0}
\begin{abstract}
We discuss about 
a possibility of large electro-weak penguin contribution in 
$B\rightarrow K \pi$ from recent experimental data. 
The several relations among the branching ratios which realize when 
the contributions from tree type and electro-weak penguin are small 
compared with the gluon penguin and can be treated as the expansion 
parameters do not satisfy the data. The difference comes from the $r^2$ 
terms which is the square of the ratio with the gluon penguin 
and the main contribution comes from electro-weak penguin. 
We find that the contribution from electro-weak penguin 
may be large to explain the experimental data. 
If the magnitude estimated from experiment is quite large compared 
with the theoretical estimation, then it may be including 
some new physics effects.      

\end{abstract}

\vspace*{6cm}

\begin{flushleft}
\hspace*{0.9cm} \begin{tabular}{l} \\ \hline 
{\small Emails: tadashi.yoshikawa@kek.jp} \\
\end{tabular}
\end{flushleft}

\end{titlepage}

One of the main targets at the B factories is to determine the all
CP angles in the unitarity triangles of the Cabbibo-Kobayashi-Maskawa (CKM)
matrix \cite{CKM}.  $\phi_1$\cite{SM} as one of the angles has already been
measured and established the CP violation in the B meson system
by Belle\cite{BF1} and BaBar\cite{BF2} collaborations.
The next step is to measure the remaining
angles.  The canonical decay modes for measuring $\phi_2$ and $\phi_3$ are
$B_d^0 \rightarrow \pi^+ \pi^-$ and $B^\pm \rightarrow D K^\pm $ respectively
but the methods have some difficulties to extract cleanly the angles. 
To avoid the difficulty, isospin relation\cite{ISO} 
and $SU(3)$ relation including 
$B\rightarrow K\pi$ modes \cite{GHLR1}-\cite{KLY} 
are being considered as a method to extract the weak phases.

$B\rightarrow K\pi $ modes has been also measured\cite{TOMURA} and they will 
be useful informations to understand the CP violation through the KM phases. 
If we can directory solve about these modes, it is very elegant way to 
determine the parameters and the weak phase. 
However we can not do so because there are 
too many parameters in $B\rightarrow K\pi $ modes 
to extract the weak phases. So we need understand these modes step by step.
To understand the weak phase through this mode, there are several 
approaches by diagram decomposition\cite{GHLR1}-\cite{FM}, 
QCD factorization\cite{BBNS} and  pQCD\cite{YLS} and so on. 
The contributions including the weak phase $\phi_3$ come from tree type 
diagrams which has CKM suppression factor and they are usually dealt with 
small parameter compared with gluon penguin. 
If we can deal the contributions except for gluon penguin with 
the small parameters, then, there are several relations among 
the averaged branching ratios of $B\rightarrow K\pi$ modes. For example, 
$Br(K^+ \pi^- )/2 Br(K^0 \pi^0 ) \approx 2 Br(K^+ \pi^0)/Br(K^0 \pi^+)$ 
\cite{BBNS} et al. However, the recent experiment does not seem to satisfy 
them so well. When we reconsider these modes to compare with the data, 
we find that the role of a color favored electro-weak penguin 
may be important to explain 
the difference between the relations and the experimental data. 
The color favored 
electro-weak penguin diagram is included in $B^0\rightarrow K^0 \pi^0$ and 
$B^0\rightarrow K^+ \pi^0$ and the data of the branching ratio are 
slightly larger than half of $B^0\rightarrow K^+ \pi^-$, where the $1/2$ 
comes from the difference of $\pi^0$ and $\pi^+$ in final state. 
So we need to know the informations about the electro-weak penguin 
contributions in $B \rightarrow K \pi$ decay modes to understand about 
the effect from the weak phases.  
The role was pointed out and the magnitude was estimated 
in several works\cite{BBNS}. They said that the ratio between 
gluon and electro-weak penguins  is about $0.14$ as the central value 
but the experimental data may suggest that the magnitude seem to be slightly 
larger than the estimation. If there is quite large deviation 
in the contribution 
from the electro-weak penguin, it may suggest a possibility of new physics in 
these modes.  

In $B^0\rightarrow K_S \pi^0$, we can consider to use the time-dependent CP 
asymmetry to extract the weak phase. However the mode has the electro-weak 
penguin diagram so that one must remove the contribution to extract $\phi_3$. 
And we have to check whether  extracting $\phi_3$ is possible or not. 
To do so, it is important to reconsider about the electro-weak penguin.    

In this work, we consider how large contribution 
from the electro-weak penguins from only experimental data.   

The decay amplitudes of $B \rightarrow K\pi$ are 
\bea
A^{0+} &\equiv & A(B^+\rightarrow K^0 \pi^+) \nn \\
         &=& 
         \left[ A V_{ub}^*V_{us} 
             + \sum_{i=u,c,t}( P_i + EP_i - \frac{1}{3}P_{EWi} 
                                  + \frac{2}{3}EP^C_{EWi} )
                       V_{ib}^*V_{is} 
                                        \right]  , \\[5mm]
A^{00} &\equiv & A(B^0\rightarrow K^0\pi^0) \nn \\
         &=& 
      - \frac{1}{\sqrt{2}}
         \left[ C  V_{ub}^*V_{us} 
             - \sum_{i=u,c,t}( P_i + EP_i - P_{EWi} - \frac{1}{3}P^C_{EWi} 
                                                    - \frac{1}{3}EP^C_{EWi})
                       V_{ib}^*V_{is} 
                                        \right]  , \\[5mm]
A^{+-} &\equiv & A(B^0\rightarrow K^+\pi^-) \nn \\
          &=&
            - \left[ T  V_{ub}^*V_{us} 
             + \sum_{i=u,c,t}( P_i + EP_i + \frac{2}{3}P^C_{EWi}
                                   - \frac{1}{3}EP^C_{EWi} )
                       V_{ib}^*V_{is} 
                                        \right]  , \\[5mm]
A^{+0} &\equiv & A(B^+\rightarrow K^+ \pi^0) \nn \\
          &=& 
            - \frac{1}{\sqrt{2}}
                      \left[ (T + C + A)  V_{ub}^*V_{us} \right.\nn \\
          &\ & ~~~~~~~~~~~~~~~~
             \left.
        + \sum_{i=u,c,t}( P_i  + EP_i + P_{EWi} + \frac{2}{3}P^C_{EWi}
                                   + \frac{2}{3}EP^C_{EWi} )
                       V_{ib}^*V_{is} 
                                        \right]  , 
\eea 
where $T$ is a color favored tree amplitude, $C$ is a Color suppressed 
tree, $A$ is an annihilation, $P_i$$(i=u,c,t)$ 
is a gluonic penguin, $EP_i$ is a penguin exchange, $P_{EWi}$ is a color 
favored electroweak penguin, $P_{EWi}^C$ is a color suppressed 
electroweak penguin and $EP_{EWi}^C$ is a color suppressed electro-weak 
penguin 
exchange diagram. After this, for simplicity, we neglect 
the u- and c- electroweak penguin diagram because of the smallness, And 
we redefine the each terms as following 
\bea
T + P_u +EP_u - P_c - EP_c &\rightarrow & T, \\
C - P_u -EP_u + P_c + EP_c &\rightarrow & C, \\
A + P_u +EP_u - P_c - EP_c &\rightarrow & A, \\
P_t + EP_t  - P_c - EP_c - \frac{1}{3}P_{EW}^C + \frac{2}{3}EP_{EW}^C 
                           &\rightarrow & P, \\ 
P_{EW}+EP_{EW}^C  &\rightarrow & P_{EW}, \\
P_{EW}^C-EP_{EW}^C  &\rightarrow & P_{EW}^C. 
\label{redP}
\eea 
One can reduce the number of parameter up to 6 (or 12). 
Then, the amplitudes are, by using the unitarity relation of KM matrix, 
\bea
A^{0+} &=& 
         \left[ P V_{tb}^*V_{ts} + A V_{ub}^*V_{us} 
                                        \right]  , \\
A^{00} &=& 
     \frac{1}{\sqrt{2}}
         \left[ (P - P_{EW} )V_{tb}^*V_{ts} 
             - C
                       V_{ub}^*V_{us} 
                                        \right]  , \\
A^{+-} &=&
            - \left[ (P + P_{EW}^C)  V_{tb}^*V_{ts} 
             + T
                       V_{ub}^*V_{us} 
                                        \right]  , \\
A^{+0} &=& 
            - \frac{1}{\sqrt{2}}
                      \left[ (P + P_{EW} + P_{EW}^C)  V_{tb}^*V_{ts} 
             + ( T + C + A )
                       V_{ub}^*V_{us} 
                                        \right]  . 
\eea 
By this diagram decomposition\cite{GHLR}, 
one can easily find the isospin relation among the amplitudes, 
\bea
\sqrt{2}A^{+0} + A^{0+} = \sqrt{2} A^{00} + A^{+-}. 
\label{isospin}
\eea 
They are rewritten as follows:
\bea
A^{0+} &=& - P |V_{tb}^*V_{ts}|
         \left[ 1 - r_A e^{i\delta^A}e^{i\phi_3}  
                                        \right]  , \\
A^{00} &=& 
     - \frac{1}{\sqrt{2}} P |V_{tb}^*V_{ts}|
         \left[ 1 - r_{EW}e^{i\delta^{EW}}  
             + r_C e^{i\delta^C}e^{i\phi_3}
                                        \right]  , \\
A^{+-} &=&
            P |V_{tb}^*V_{ts}| 
            \left[ 1 + r_{EW}^C  e^{i\delta^{EWC}} 
             - r_T e^{i\delta^T}e^{i\phi_3}
                                        \right]  , \\
A^{+0} &=&  \frac{1}{\sqrt{2}}
            P |V_{tb}^*V_{ts}|
                      \left[ 1 + r_{EW} e^{i\delta^{EW}} 
                               + r_{EW}^C e^{i\delta^{EWC}} 
             - ( r_Te^{i\delta^T} + r_C e^{i\delta^C} + r_A e^{i\delta^A} )
                       e^{i\phi_3} 
                                        \right], 
\eea 
where
$\phi_3$ is the weak phase of $V_{ub}^*V_{us}$, $\delta^X$ is 
the strong phase difference between each diagram and gluon penguin and 
\bea 
r_A &=& \frac{ |A V_{ub}^*V_{us}| }{ |P V_{tb}^*V_{ts}|}, ~~~~
r_T = \frac{ |T V_{ub}^*V_{us}| }{ |P V_{tb}^*V_{ts}|}, ~~~~ 
r_C = \frac{ |C V_{ub}^*V_{us}| }{ |P V_{tb}^*V_{ts}|}, \\
& & r_{EW} = \frac{ |P_{EW}|}{ |P|}, ~~~~
r_{EW}^C = \frac{ |P_{EW}^C|}{ |P|}.
\eea
We assume as the hierarchy of the ratios that 
$ 1 > r_T, r_{EW} > r_C, r_{EW}^C > r_A $ ~\cite{GHLR}. 
$|P/T|$ was estimated about $0.1$ in \cite{YOSHI}
\footnote{ Note that this ratio $|P/T|$ does not include CKM factors. } 
by considering the 
$B\rightarrow \pi \pi$ and it was also shown by the ratio of branchings 
of $B^+ \rightarrow \pi^0 \pi^+$ and $B^+ \rightarrow K^0 \pi^+$
\cite{LU,CONV1,CONV2}.  In $B\rightarrow K\pi$ mode, the tree type diagram 
is suppressed by KM factor $V_{ub}^*V_{us}$ and 
$r_T \sim |T/P| \times \lambda^2 R_b \sim 0.2 $, where cabbibo angle 
$\lambda = 0.22$ and we used $R_b = \sqrt{\rho^2+\eta^2} \sim 0.4 $. 
$r_C$ and $r_{EW}^C$ are suppressed by color factor from $r_T$ and $r_{EW}$. 
Comparing the Wilson coefficients which correspond to the diagrams 
under the factorization method, 
we assume that  $r_C \sim 0.1 r_{T}$ and $r_{EW}^C \sim 0.1 r_{EW}$
\cite{LU,BBNS}. Here we do not put any assumption about the magnitude of 
$r_{EW}$.  $r_A$ could be negligible because it should have B meson 
decay constant and it works as a suppression factor $f_B/M_B$.   
According to this assumption, 
we will neglect the $r^2$ 
terms including $r_C, r_A$ and $r_{EW}^C$.     
Then, the averaged branching ratios are 
\bea
\bar{B}^{0+}&\propto & \frac{1}{2}\left[|A^{0+}|^2 + |A^{0-}|^2\right] \nn \\
       &=& |P|^2 |V_{tb}^*V_{ts}|^2 
             \left[ 1 - 2 r_A \cos\delta^A\cos\phi_3  \right],
\label{B0+} \\[5mm]
2 \bar{B}^{00}&\propto &   
                           \left[|A^{00}|^2 + |\bar{A}^{00}|^2\right]
      \nn \\ 
      &=& |P|^2 |V_{tb}^*V_{ts}|^2 
                \left[ 1 + r_{EW}^2 
                       - 2 r_{EW} \cos\delta^{EW} 
                       + 2 r_C \cos\delta^C\cos\phi_3
                \right], \\[5mm]
\bar{B}^{+-}&\propto & \frac{1}{2}\left[|A^{+-}|^2 + |A^{-+}|^2\right]
       \nn \\
       &=& |P|^2 |V_{tb}^*V_{ts}|^2 
                \left[ 1 + r_T^2 
                       + 2 r_{EW}^C \cos\delta^{EWC} 
                       - 2 r_T \cos\delta^T\cos\phi_3
                \right], \\[5mm]
2 \bar{B}^{+0}&\propto & \left[|A^{+0}|^2 + |A^{-0}|^2\right]
       \nn \\
       &=& |P|^2 |V_{tb}^*V_{ts}|^2 
                \left[ 1 +  r_{EW}^2 + r_T^2 
                       + 2 r_{EW} \cos\delta^{EW} 
                       + 2 r_{EW}^C \cos\delta^{EWC} \right.\nn \\
       & & ~~~~~~~~~~
                       - ( 2 r_T \cos\delta^T + 2 r_C \cos\delta^C 
                         + 2 r_A \cos\delta^A )\cos\phi_3 \nn \\
       & & ~~~~~~~~~~  \left.
                - 2 r_{EW} r_T \cos(\delta^{EW}-\delta^{T})\cos\phi_3
                \right]. 
\label{B+0}
\eea

\begin{table}
\begin{center}
\begin{tabular}{|c|c|c|c|c|}\hline
  & CLEO\cite{KPCLEO}
  & Belle\cite{TOMURA,KPBELLE} & BaBar\cite{KPBABAR1,KPBABAR2} & Average \\
\hline
$Br(B^+ \rightarrow K^0 \pi^+) \times 10^{6} $ 
             & 18.8 ${}^{+3.7+2.1}_{-3.3-1.8}$
                          & 22.0 $\pm$ 1.9
                          $\pm$ 1.1
                          & 17.5 ${}^{+1.8}_{-1.7}$ $\pm$ 1.3
                          & 19.6 $\pm$ 1.4  \\[2mm]
$Br(B^0 \rightarrow K^0 \pi^0) \times 10^{6} $ 
             & 12.8 ${}^{+4.0+1.7}_{-3.3-1.4}$
                          & 12.6 $\pm$ 2.4
                          $\pm$ 1.4
                          & 10.4 $\pm$1.5 $\pm$ 0.8
                          & 11.2 $\pm$ 1.4  \\[2mm]
$Br(B^0 \rightarrow K^+ \pi^-) \times 10^{6} $ 
             & 18.0${}^{+2.3+1.2}_{-2.1-0.9}$ 
                          & 18.5 $\pm$1.0
                          $\pm$ 0.7
                          & 17.9 $\pm$ 0.9 $\pm$ 0.7
                          & 18.2 $\pm$ 0.8  \\[2mm]
$Br(B^+ \rightarrow K^+ \pi^0) \times 10^{6} $ 
             & 12.9${}^{+2.4+1.2}_{-2.2-1.1}$ 
                          & 12.8 $\pm$1.4
                          ${}^{+1.4}_{-1.0}$ 
                          & 12.8 ${}^{+1.2}_{-1.1}$ $\pm$ 1.0
                          & 12.8 $\pm$ 1.1  \\
\hline
\end{tabular}
\caption{The experimental data and the average. }
\end{center}
\end{table}

One can take several ratios between the branching ratios. If all modes are 
gluon penguin dominant, the ratios should be close to $1$. 
The shift from $1$ will depend on the magnitude of $r$s.   
From the averaged values of the recent experimental data in Table 1,  
\bea
\frac{\bar{B}^{+-}}{2\bar{B}^{00}} &=& 0.81 \pm 0.11, 
\label{data+-}\\
\frac{2 \bar{B}^{+0}}{\bar{B}^{0+}} &=& 1.31 \pm 0.15, \\[5mm]
\frac{\tau^+}{\tau^0}\frac{\bar{B}^{+-}}{\bar{B}^{0+}} &=& 1.01 \pm 0.09, \\
\frac{\tau^0}{\tau^+}\frac{\bar{B}^{+0}}{\bar{B}^{00}} &=& 1.05 \pm 0.16, 
                                                     \\[5mm]
\frac{\tau^+}{\tau^0}\frac{2 \bar{B}^{00}}{\bar{B}^{0+}} &=& 1.24 \pm 0.18, \\
\frac{\tau^0}{\tau^+}\frac{2 \bar{B}^{+0}}{\bar{B}^{+-}} &=& 1.30 \pm 0.13,
\label{data+0}  
\eea 
where $\frac{\tau^+}{\tau^0}$ is a lifetime ratio 
between the charged and the neutral $B$ mesons and 
$\tau(B^\pm)/\tau(B^0) = 1.083 \pm 0.017$\cite{PDG}. 
While, from eq.(\ref{B0+})-(\ref{B+0}) under the assumption that 
all $r$ is smaller than $1$ and the $r^2$ terms including $r_C, r_A$ 
and $r_{EW}^C$ are neglected,         
\bea
\frac{\bar{B}^{+-}}{2\bar{B}^{00}} &=& \left\{
  1 + 2 r_{EW} \cos\delta^{EW} + 2 r_{EW}^C \cos\delta^{EWC} 
    - 2 ( r_{T} \cos\delta^{T} + r_C \cos\delta^C )\cos\phi_3  
    + r_{T}^2 \right\}\nn \\
  & & ~~~ -r_{EW}^2 + 4 r_{EW}^2 \cos^2\delta^{EW}, 
\label{B+-B00}\\[4mm]
\frac{2 \bar{B}^{+0}}{\bar{B}^{0+}} &=&
 \left\{ 1 + 2 r_{EW} \cos\delta^{EW} + 2 r_{EW}^C \cos\delta^{EWC} 
    - 2 ( r_{T} \cos\delta^{T} + r_C \cos\delta^C )\cos\phi_3  
  + r_{T}^2 \right\} \nn \\
  & & ~~~ + r_{EW}^2 
            - 2 r_{EW} r_T \cos(\delta^{EW}-\delta^T)\cos\phi_3, 
\label{B+0B0+}\\[4mm]
\frac{\tau^+}{\tau^0}\frac{\bar{B}^{+-}}{\bar{B}^{0+}} &=&
  1 + 2 r_{EW}^C \cos\delta^{EWC} 
    - 2 ( r_{T} \cos\delta^{T} - r_A \cos\delta^A )\cos\phi_3  
    + r_{T}^2,  \label{B+-B0+}\\[4mm]  
\frac{\tau^0}{\tau^+}\frac{\bar{B}^{+0}}{\bar{B}^{00}} &=& 
 1 + 2 r_{EW}^C \cos\delta^{EWC} 
    - 2 ( r_{T} \cos\delta^{T} 
          + 2 r_C \cos\delta^C + r_A \cos\delta^A )\cos\phi_3 + r_{T}^2 \nn \\
  & & ~~~ + 4 r_{EW} \cos\delta^{EW} 
        - 2 r_{EW} r_T \cos(\delta^{EW}-\delta^T)\cos\phi_3
  + 4 r_{EW}^2 \cos^2\delta^{EW}, \label{B+0B00} \\[4mm]
\frac{\tau^+}{\tau^0}\frac{2 \bar{B}^{00}}{\bar{B}^{0+}} &=&
   1 - 2 r_{EW} \cos\delta^{EW} 
       - 2 ( r_C \cos\delta^C + r_A \cos\delta^A )\cos\phi_3 + r_{EW}^2. 
\label{B00B0+} \\[4mm]
\frac{\tau^0}{\tau^+}\frac{2 \bar{B}^{+0}}{\bar{B}^{+-}} &=&
   1 + 2 r_{EW} \cos\delta^{EW} 
       - 2 ( r_C \cos\delta^C + r_A \cos\delta^A )\cos\phi_3 + r_{EW}^2
                               \nn \\
   & & ~~~ 
       - 2 r_{EW} r_T \cos(\delta^{EW}-\delta^T)\cos\phi_3
       +4r_T^2 \cos^2\delta^T \cos^2\phi_3. 
\label{B+0B+-}
\eea
If all modes are dominated by only gluonic penguin contribution, namely, 
all $r$ is much smaller than $1$, then all ratio of branching ratios should
be $1$. But the data are not so. eqs.(\ref{B+-B00}),(\ref{B+0B0+}),
(\ref{B00B0+}) and (\ref{B+0B+-}) seem to differ from $1$ so that 
there must be some sizable contributions except for the gluon penguin.   

\begin{figure}[htbp]
\begin{center}
\begin{minipage}[l]{3.0in}
\setlength{\unitlength}{0.080450pt}
\begin{picture}(2699,2069)(0,0)
\footnotesize
\thicklines \path(370,249)(411,249)
\thicklines \path(2576,249)(2535,249)
\put(329,249){\makebox(0,0)[r]{ 0}}
\thicklines \path(370,539)(411,539)
\thicklines \path(2576,539)(2535,539)
\put(329,539){\makebox(0,0)[r]{ 0.2}}
\thicklines \path(370,829)(411,829)
\thicklines \path(2576,829)(2535,829)
\put(329,829){\makebox(0,0)[r]{ 0.4}}
\thicklines \path(370,1119)(411,1119)
\thicklines \path(2576,1119)(2535,1119)
\put(329,1119){\makebox(0,0)[r]{ 0.6}}
\thicklines \path(370,1408)(411,1408)
\thicklines \path(2576,1408)(2535,1408)
\put(329,1408){\makebox(0,0)[r]{ 0.8}}
\thicklines \path(370,1698)(411,1698)
\thicklines \path(2576,1698)(2535,1698)
\put(329,1698){\makebox(0,0)[r]{ 1}}
\thicklines \path(370,1988)(411,1988)
\thicklines \path(2576,1988)(2535,1988)
\put(329,1988){\makebox(0,0)[r]{ 1.2}}
\thicklines \path(370,249)(370,290)
\thicklines \path(370,1988)(370,1947)
\put(370,166){\makebox(0,0){-1}}
\thicklines \path(922,249)(922,290)
\thicklines \path(922,1988)(922,1947)
\put(922,166){\makebox(0,0){-0.5}}
\thicklines \path(1473,249)(1473,290)
\thicklines \path(1473,1988)(1473,1947)
\put(1473,166){\makebox(0,0){ 0}}
\thicklines \path(2025,249)(2025,290)
\thicklines \path(2025,1988)(2025,1947)
\put(2025,166){\makebox(0,0){ 0.5}}
\thicklines \path(2576,249)(2576,290)
\thicklines \path(2576,1988)(2576,1947)
\put(2576,166){\makebox(0,0){ 1}}
\thicklines \path(370,249)(2576,249)(2576,1988)(370,1988)(370,249)
\put(22,1218){\makebox(0,0)[l]{$r_{EW}$}}
\put(1473,42){\makebox(0,0){ $ \cos\delta^{EW} $}}
\thinlines \path(1837,1916)(1837,1916)(1843,1908)(1848,1901)
(1854,1894)(1859,1887)(1859,1879)(1859,1872)(1865,1865)(1865,1858)
(1865,1850)(1865,1843)(1870,1836)(1870,1829)(1870,1821)(1870,1814)
(1870,1807)(1870,1800)(1870,1792)(1870,1785)(1870,1778)(1870,1771)
(1870,1763)(1870,1756)(1876,1749)(1876,1742)(1876,1734)(1876,1727)
(1876,1720)(1876,1713)(1870,1705)(1870,1698)(1870,1691)(1870,1684)
(1870,1676)(1870,1669)(1870,1662)(1870,1655)(1870,1647)(1870,1640)
(1870,1633)(1870,1626)(1870,1618)(1870,1611)(1870,1604)(1865,1597)
(1865,1589)(1865,1582)(1865,1575)(1865,1568)(1865,1560)
\thinlines \path(1865,1560)(1865,1553)(1865,1546)(1865,1539)(1859,1532)
(1859,1524)(1859,1517)(1859,1510)(1859,1503)(1859,1495)(1854,1488)
(1854,1481)(1854,1474)(1854,1466)(1854,1459)(1854,1452)(1848,1445)
(1848,1437)(1848,1430)(1848,1423)(1848,1416)(1843,1408)(1843,1401)
(1843,1394)(1843,1387)(1843,1379)(1837,1372)(1837,1365)(1837,1358)
(1837,1350)(1837,1343)(1831,1336)(1831,1329)(1831,1321)(1831,1314)
(1826,1307)(1826,1300)(1826,1292)(1826,1285)(1820,1278)(1820,1271)
(1820,1263)(1820,1256)(1815,1249)(1815,1242)(1815,1234)(1809,1227)
(1809,1220)(1809,1213)(1809,1205)(1804,1198)
\thinlines \path(1804,1198)(1804,1191)(1804,1184)(1798,1176)(1798,1169)
(1798,1162)(1798,1155)(1793,1147)(1793,1140)(1793,1133)(1787,1126)
(1787,1119)(1787,1111)(1782,1104)(1782,1097)(1782,1090)(1776,1082)
(1776,1075)(1776,1068)(1771,1061)(1771,1053)(1771,1046)(1765,1039)
(1765,1032)(1765,1024)(1760,1017)(1760,1010)(1760,1003)(1754,995)
(1754,988)(1754,981)(1749,974)(1749,966)(1749,959)(1743,952)(1743,945)
(1743,937)(1738,930)(1738,923)(1738,916)(1732,908)(1732,901)(1732,894)
(1727,887)(1727,879)(1721,872)(1721,865)(1721,858)(1716,850)(1716,843)
(1716,836)
\thinlines \path(1716,836)(1710,829)(1710,821)(1710,814)(1705,807)
(1705,800)(1705,792)(1699,785)(1699,778)(1699,771)(1694,763)(1694,756)
(1694,749)(1688,742)(1688,734)(1683,727)(1683,720)(1683,713)(1677,705)
(1677,698)(1677,691)(1677,684)(1672,677)(1672,669)(1672,662)(1666,655)
(1666,648)(1666,640)(1661,633)(1661,626)(1661,619)(1661,611)(1655,604)
(1655,597)(1655,590)(1649,582)(1649,575)(1649,568)(1649,561)(1649,553)
(1644,546)(1644,539)(1644,532)(1644,524)(1644,517)(1644,510)(1644,503)
(1644,495)(1638,488)(1638,481)(1638,474)
\thinlines \path(1638,474)(1638,466)(1644,459)(1644,452)(1644,445)
(1638,437)(1627,430)(1611,423)(1589,416)(1550,408)(1495,401)(1412,394)
(1274,387)
\thinlines \path(1837,1916)(1837,1916)(1826,1908)(1820,1901)(1809,1894)
(1804,1887)(1798,1879)(1793,1872)(1787,1865)(1787,1858)(1782,1850)
(1776,1843)(1771,1836)(1771,1829)(1765,1821)(1760,1814)(1760,1807)
(1754,1800)(1749,1792)(1749,1785)(1743,1778)(1738,1771)(1738,1763)
(1732,1756)(1727,1749)(1727,1742)(1721,1734)(1716,1727)(1716,1720)
(1710,1713)(1705,1705)(1705,1698)(1699,1691)(1699,1684)(1694,1676)
(1688,1669)(1688,1662)(1683,1655)(1677,1647)(1677,1640)(1672,1633)
(1666,1626)(1666,1618)(1661,1611)(1655,1604)(1655,1597)(1649,1589)
(1644,1582)(1638,1575)(1638,1568)(1633,1560)
\thinlines \path(1633,1560)(1627,1553)(1627,1546)(1622,1539)(1616,1532)
(1611,1524)(1611,1517)(1605,1510)(1600,1503)(1594,1495)(1594,1488)
(1589,1481)(1583,1474)(1578,1466)(1578,1459)(1572,1452)(1567,1445)
(1561,1437)(1556,1430)(1550,1423)(1550,1416)(1545,1408)(1539,1401)
(1534,1394)(1528,1387)(1523,1379)(1517,1372)(1512,1365)(1506,1358)
(1501,1350)(1501,1343)(1495,1336)(1490,1329)(1484,1321)(1479,1314)
(1473,1307)(1467,1300)(1462,1292)(1456,1285)(1445,1278)(1440,1271)
(1434,1263)(1429,1256)(1423,1249)(1418,1242)(1412,1234)(1407,1227)
(1401,1220)(1390,1213)(1385,1205)(1379,1198)
\thinlines \path(1379,1198)(1374,1191)(1368,1184)(1357,1176)(1352,1169)
(1346,1162)(1335,1155)(1330,1147)(1324,1140)(1313,1133)(1308,1126)
(1302,1119)(1291,1111)(1285,1104)(1274,1097)(1269,1090)(1258,1082)
(1252,1075)(1241,1068)(1236,1061)(1225,1053)(1219,1046)(1208,1039)
(1197,1032)(1192,1024)(1181,1017)(1170,1010)(1164,1003)(1153,995)
(1142,988)(1131,981)(1120,974)(1109,966)(1103,959)(1092,952)(1081,945)
(1070,937)(1059,930)(1048,923)(1037,916)(1026,908)(1010,901)(999,894)
(988,887)(977,879)(966,872)(949,865)(938,858)(922,850)(910,843)(899,836)
\thinlines \path(899,836)(883,829)(866,821)(855,814)(839,807)(822,800)
(811,792)(795,785)(778,778)(762,771)(745,763)(728,756)(712,749)(695,742)
(673,734)(657,727)(635,720)(618,713)(596,705)(580,698)(558,691)(535,684)
(513,677)(491,669)(464,662)(442,655)(414,648)(392,640)(370,633)(370,633)
\thinlines \path(370,401)(458,394)(602,387)(1274,387)
\end{picture}

\end{minipage}
    \hspace*{8mm}
\begin{minipage}[r]{3.0in}
\setlength{\unitlength}{0.080450pt}
\begin{picture}(2699,2069)(0,0)
\footnotesize
\thicklines \path(370,249)(411,249)
\thicklines \path(2576,249)(2535,249)
\put(329,249){\makebox(0,0)[r]{ 0}}
\thicklines \path(370,539)(411,539)
\thicklines \path(2576,539)(2535,539)
\put(329,539){\makebox(0,0)[r]{ 0.2}}
\thicklines \path(370,829)(411,829)
\thicklines \path(2576,829)(2535,829)
\put(329,829){\makebox(0,0)[r]{ 0.4}}
\thicklines \path(370,1119)(411,1119)
\thicklines \path(2576,1119)(2535,1119)
\put(329,1119){\makebox(0,0)[r]{ 0.6}}
\thicklines \path(370,1408)(411,1408)
\thicklines \path(2576,1408)(2535,1408)
\put(329,1408){\makebox(0,0)[r]{ 0.8}}
\thicklines \path(370,1698)(411,1698)
\thicklines \path(2576,1698)(2535,1698)
\put(329,1698){\makebox(0,0)[r]{ 1}}
\thicklines \path(370,1988)(411,1988)
\thicklines \path(2576,1988)(2535,1988)
\put(329,1988){\makebox(0,0)[r]{ 1.2}}
\thicklines \path(370,249)(370,290)
\thicklines \path(370,1988)(370,1947)
\put(370,166){\makebox(0,0){-1}}
\thicklines \path(922,249)(922,290)
\thicklines \path(922,1988)(922,1947)
\put(922,166){\makebox(0,0){-0.5}}
\thicklines \path(1473,249)(1473,290)
\thicklines \path(1473,1988)(1473,1947)
\put(1473,166){\makebox(0,0){ 0}}
\thicklines \path(2025,249)(2025,290)
\thicklines \path(2025,1988)(2025,1947)
\put(2025,166){\makebox(0,0){ 0.5}}
\thicklines \path(2576,249)(2576,290)
\thicklines \path(2576,1988)(2576,1947)
\put(2576,166){\makebox(0,0){ 1}}
\thicklines \path(370,249)(2576,249)(2576,1988)(370,1988)(370,249)
\put(22,1218){\makebox(0,0)[l]{$r_{EW}$}}
\put(1473,42){\makebox(0,0){ $ 
         r_{T} \cos(\delta^{EW}-\delta^{T})\cos \phi_3 $}}
\thinlines \path(2300,1908)(2300,1908)(2306,1901)(2311,1894)
(2311,1887)(2311,1879)(2317,1872)(2317,1865)(2311,1858)(2311,1850)
(2311,1843)(2311,1836)(2311,1829)(2311,1821)(2306,1814)(2306,1807)
(2306,1800)(2300,1792)(2300,1785)(2300,1778)(2295,1771)(2295,1763)
(2289,1756)(2289,1749)(2289,1742)(2284,1734)(2284,1727)(2278,1720)
(2278,1713)(2273,1705)(2273,1698)(2267,1691)(2262,1684)(2262,1676)
(2256,1669)(2256,1662)(2251,1655)(2251,1647)(2245,1640)(2240,1633)
(2240,1626)(2234,1618)(2229,1611)(2229,1604)(2223,1597)(2218,1589)
(2218,1582)(2212,1575)(2206,1568)(2206,1560)(2201,1553)
\thinlines \path(2201,1553)(2195,1546)(2195,1539)(2190,1532)(2184,1524)
(2179,1517)(2179,1510)(2173,1503)(2168,1495)(2162,1488)(2162,1481)
(2157,1474)(2151,1466)(2146,1459)(2140,1452)(2140,1445)(2135,1437)
(2129,1430)(2124,1423)(2118,1416)(2113,1408)(2113,1401)(2107,1394)
(2102,1387)(2096,1379)(2091,1372)(2085,1365)(2080,1358)(2074,1350)
(2069,1343)(2069,1336)(2063,1329)(2058,1321)(2052,1314)(2047,1307)
(2041,1300)(2036,1292)(2030,1285)(2025,1278)(2019,1271)(2013,1263)
(2008,1256)(2002,1249)(1997,1242)(1991,1234)(1986,1227)(1980,1220)
(1975,1213)(1969,1205)(1964,1198)(1958,1191)
\thinlines \path(1958,1191)(1953,1184)(1947,1176)(1936,1169)(1931,1162)
(1925,1155)(1920,1147)(1914,1140)(1909,1133)(1903,1126)(1898,1119)
(1887,1111)(1881,1104)(1876,1097)(1870,1090)(1865,1082)(1859,1075)
(1848,1068)(1843,1061)(1837,1053)(1831,1046)(1820,1039)(1815,1032)
(1809,1024)(1804,1017)(1793,1010)(1787,1003)(1782,995)(1771,988)
(1765,981)(1760,974)(1749,966)(1743,959)(1732,952)(1727,945)(1721,937)
(1710,930)(1705,923)(1694,916)(1688,908)(1677,901)(1672,894)(1661,887)
(1655,879)(1644,872)(1638,865)(1627,858)(1616,850)(1611,843)(1600,836)
(1589,829)
\thinlines \path(1589,829)(1583,821)(1572,814)(1561,807)(1556,800)
(1545,792)(1534,785)(1523,778)(1512,771)(1501,763)(1490,756)(1484,749)
(1473,742)(1462,734)(1445,727)(1434,720)(1423,713)(1412,705)(1401,698)
(1390,691)(1374,684)(1363,677)(1352,669)(1335,662)(1324,655)(1308,648)
(1297,640)(1280,633)(1263,626)(1247,619)(1236,611)(1219,604)(1203,597)
(1181,590)(1164,582)(1148,575)(1126,568)(1109,561)(1087,553)(1065,546)
(1048,539)(1021,532)(999,524)(977,517)(949,510)(922,503)(894,495)(866,488)
(839,481)(806,474)(773,466)
\thinlines \path(773,466)(734,459)(695,452)(657,445)(613,437)(569,430)
(519,423)(464,416)(403,408)(370,404)
\thinlines \path(2300,1908)(2273,1908)(2273,1908)(2256,1901)
(2245,1894)(2234,1887)
(2223,1879)(2218,1872)(2206,1865)(2201,1858)(2190,1850)(2184,1843)
(2173,1836)(2168,1829)(2162,1821)(2151,1814)(2146,1807)(2140,1800)
(2135,1792)(2124,1785)(2118,1778)(2113,1771)(2107,1763)(2102,1756)
(2091,1749)(2085,1742)(2080,1734)(2074,1727)(2069,1720)(2063,1713)
(2052,1705)(2047,1698)(2041,1691)(2036,1684)(2030,1676)(2025,1669)
(2013,1662)(2008,1655)(2002,1647)(1997,1640)(1991,1633)(1986,1626)
(1975,1618)(1969,1611)(1964,1604)(1958,1597)(1953,1589)(1947,1582)
(1936,1575)(1931,1568)(1925,1560)(1920,1553)
\thinlines \path(1920,1553)(1914,1546)(1909,1539)(1898,1532)(1892,1524)
(1887,1517)(1881,1510)(1870,1503)(1865,1495)(1859,1488)(1854,1481)
(1843,1474)(1837,1466)(1831,1459)(1826,1452)(1815,1445)(1809,1437)
(1804,1430)(1793,1423)(1787,1416)(1782,1408)(1771,1401)(1765,1394)
(1754,1387)(1749,1379)(1743,1372)(1732,1365)(1727,1358)(1716,1350)
(1710,1343)(1699,1336)(1694,1329)(1683,1321)(1677,1314)(1666,1307)
(1661,1300)(1649,1292)(1638,1285)(1633,1278)(1622,1271)(1611,1263)
(1605,1256)(1594,1249)(1583,1242)(1578,1234)(1567,1227)(1556,1220)
(1545,1213)(1534,1205)(1528,1198)(1517,1191)
\thinlines \path(1517,1191)(1506,1184)(1495,1176)(1484,1169)(1473,1162)
(1462,1155)(1451,1147)(1440,1140)(1429,1133)(1412,1126)(1401,1119)
(1390,1111)(1379,1104)(1368,1097)(1352,1090)(1341,1082)(1330,1075)
(1313,1068)(1302,1061)(1285,1053)(1274,1046)(1258,1039)(1247,1032)
(1230,1024)(1214,1017)(1203,1010)(1186,1003)(1170,995)(1153,988)
(1137,981)(1120,974)(1103,966)(1087,959)(1070,952)(1054,945)(1037,937)
(1021,930)(999,923)(982,916)(966,908)(944,901)(927,894)(905,887)
(888,879)(866,872)(844,865)(822,858)(800,850)(778,843)(756,836)(734,829)
\thinlines \path(734,829)(712,821)(690,814)(662,807)(640,800)(613,792)
(591,785)(563,778)(535,771)(508,763)(480,756)(453,749)(425,742)
(392,734)(370,729)
\end{picture}

\end{minipage}
\caption{The allowed region on $(r_{EW},  ~\cos \delta^{EW})$ and 
$(r_{EW}, ~r_T \cos(\delta^{EW}~-~\delta^{T}) \cos \phi_3)$ plane
at $1\sigma $ level. }
    \label{fig:1}
\end{center}
\end{figure}
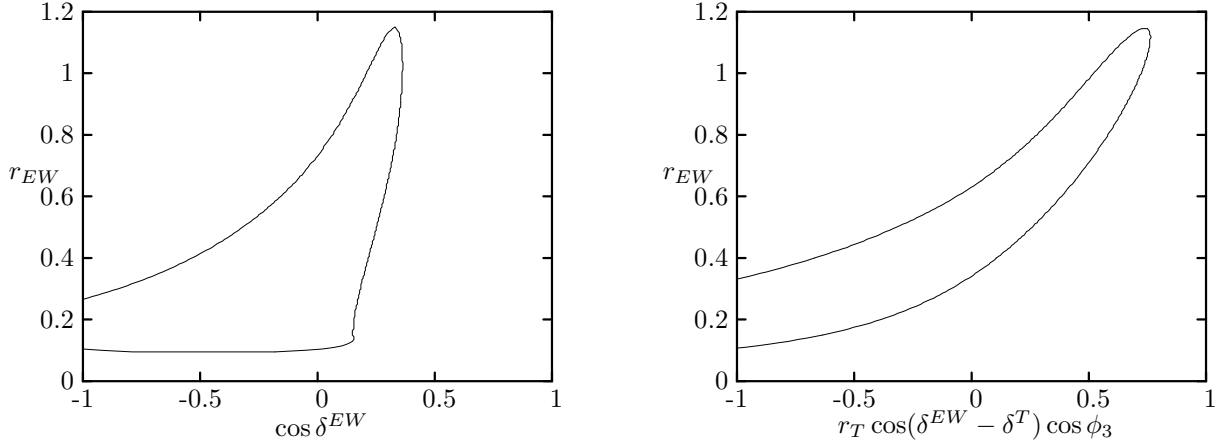

If we can neglect all $r^2$ terms, then there are a few relations among 
eqs.(\ref{B+-B00})-(\ref{B+0B+-}) as following 
\bea
\frac{\bar{B}^{+-}}{2\bar{B}^{00}} &=& \frac{2 \bar{B}^{+0}}{\bar{B}^{0+}} ,
                                   \\[3mm]
\frac{2 \bar{B}^{+0}}{\bar{B}^{0+}}
 - \frac{\tau^+}{\tau^0}\frac{\bar{B}^{+-}}{\bar{B}^{0+}} &+& 
\frac{\tau^+}{\tau^0}\frac{2 \bar{B}^{00}}{\bar{B}^{0+}} -1 = 0 . 
\label{BISO}
\eea 
However, the experimental data listed in eqs.(\ref{data+-})-(\ref{data+0})
do not satisfy these relations so well. According 
to the experimental data, 
$\frac{\bar{B}^{+-}}{2\bar{B}^{00}}$ seems to be smaller than 1 but 
$\frac{2 \bar{B}^{+0}}{\bar{B}^{0+}}$ be larger than 1. So it shows 
there is a discrepancy between them.  
The equations of 
$\frac{\bar{B}^{+-}}{2\bar{B}^{00}}$ and 
$\frac{2 \bar{B}^{+0}}{\bar{B}^{0+}}$ are same up to $r_T^2$ term 
and the difference comes from $r^2$ term including $r_{EW}$. 
$r_T^2$ term does not seem to contribute to the ratios so strongly. 
The second relation corresponds to the isospin relation of 
eq.(\ref{isospin}) at the first order of $r$.  
The discrepancy of relation (\ref{BISO}) from $0$ also comes from 
$r^2$ term including $r_{EW}$. 
The differences are 
\bea
\frac{2 \bar{B}^{+0}}{\bar{B}^{0+}} - \frac{\bar{B}^{+-}}{2\bar{B}^{00}} = 
   2 r_{EW}^2 - 2 r_{EW} r_{T} \cos(\delta^{EW}-\delta^{T})\cos\phi_3
              - 4 r_{EW}^2 \cos^2\delta^{EW} = 0.50 \pm 0.19 ,
\label{B+0B0+MB+-B00} \\[3mm]
\frac{2 \bar{B}^{+0}}{\bar{B}^{0+}}
 - \frac{\tau^+}{\tau^0}\frac{\bar{B}^{+-}}{\bar{B}^{0+}} + 
\frac{\tau^+}{\tau^0}\frac{2 \bar{B}^{00}}{\bar{B}^{0+}} -1 = 
2 r_{EW}^2 - 2 r_{EW} r_{T} \cos(\delta^{EW}-\delta^{T})\cos\phi_3 
                                                      = 0.54 \pm 0.25 , 
\label{B+0MB+-PB00M1}
\eea 
so that one can find the electro-weak penguin contributions may be large. 
All terms are including $r_{EW}$ and the deviation of the relation from $0$
is finite. Here the error are determined by adding quadratically all errors. 
Using the other relation as following  
\bea
 & & \frac{\bar{B}^{+-}}{2\bar{B}^{00}}
 - \frac{\tau^0}{\tau^+}\frac{\bar{B}^{+0}}{\bar{B}^{00}} + 
\frac{\tau^+}{\tau^0}\frac{2 \bar{B}^{00}}{\bar{B}^{0+}} - 1 \nn \\
& & ~~~~~~~~~~~~~~~~~~~~~~~  = 
- 4 r_{EW} \cos\delta^C 
          + 2 r_{EW} r_{T} \cos(\delta^{EW}-\delta^{T})\cos\phi_3 
                                                      = 0.00 \pm 0.26 , 
\label{B+-MB+0PB00M1}
\eea 
we can solve them about $r_{EW}^2$ and if we can respect the central values,
the solutions are
\bea
 & & (r_{EW},  ~\cos\delta^{EW}, 
   ~r_{T} \cos(\delta^{EW}-\delta^{T})\cos\phi_3 ) \nn \\
& & ~~~~~~~~~~~~~~~~~~~~~~~~~~~~~~~~~~~~~~~~  = 
( 0.26,  -0.38, -0.75 ) ~\mbox{and} ~( 0.69, 0.21, 0.41 ) . 
\label{solution}
\eea
This solution show that large electro-weak penguin contribution (but 
$r_{T} \cos(\delta^{EW}-\delta^{T})\cos\phi_3$ is too large because 
$r_T$ was estimated around 0.2 by the other methods.)   
The allowed region of $r_{EW}$, 
$\cos\delta^{EW}$ and $r_{T} \cos(\delta^{EW}-\delta^{T})\cos\phi_3 $ at 
$1 \sigma $ level for eqs.(\ref{B+0B0+MB+-B00})-(\ref{B+-MB+0PB00M1}) 
is shown in Fig.\ref{fig:1}. 

From this result, we find that the smaller $r_{EW}$ will favored 
a larger $|r_T \cos(\delta^{EW}~-~\delta^{T}) \cos \phi_3|$ term 
with negative sign. 
However such large $r_{T}$ is disfavored by the rough estimation $r_T$ 
should be around 0.2 which will satisfy that $|P/T|\sim 0.1$ to explain
the large CP violation in $B^0\rightarrow \pi^+\pi^-$.  
Even if $|r_T \cos(\delta^{EW}~-~\delta^{T}) \cos \phi_3|$ is within 0.2, 
then $r_{EW}$ will also be larger than 0.2 and $r_{EW}$ will be larger than
$r_T$.  This is showing that there is 
a possibility of large electro-weak penguin contribution.
Note that in the case $r_{EW}$
is quite large, the expansion by $r_{EW}$ may not be good but 
eq.(\ref{B+0MB+-PB00M1}) will be satisfied. 
Roughly speaking, the shift of eqs.(\ref{data+-})-(\ref{data+0}) from $1$ 
seem to depend on the $r_{EW}^2$ term and the sign. 
To fix the solution or confirm the large electro-weak penguin contribution, 
we need higher accurate data.      

The contributions from tree diagram are not so large except for the cross 
term with the electro-weak penguin because 
$\frac{\tau^+}{\tau^0}\frac{\bar{B}^{+-}}{\bar{B}^{0+}}$ is quite near $1$. 
When we consider the direct CP asymmetry, the experimental data in Table. 2 
do not also show so large value.    
\begin{table}
\begin{center}
\begin{tabular}{|c|c|c|c|}\hline
  & Belle\cite{TOMURA,AKPBELLE2} & BaBar\cite{KPBABAR1,KPBABAR2} & Average \\
\hline
$A_{CP}(B^+ \rightarrow K^0 \pi^+) $ 
             & 0.07 ${}^{+0.09 + 0.01}_{-0.08 - 0.03}$
                          & -0.17 $\pm$ 0.10
                          $\pm$ 0.02
                          & -0.03 $\pm$ 0.07 \\[2mm]
$A_{CP}(B^0 \rightarrow K^0 \pi^0)  $ 
             & -
                          & 0.03 $\pm$ 0.036
                          $\pm$ 0.09
                          & 0.03 $\pm$ 0.37  \\[2mm]
$A_{CP}(B^0 \rightarrow K^+ \pi^-) $ 
             & -0.07 $\pm$ 0.06 $\pm$ 0.01 
                          & -0.10
                          $\pm$ 0.05 $\pm $ 0.02 
                          & -0.09 $\pm$ 0.04  \\[2mm]
$A_{CP}(B^+ \rightarrow K^+ \pi^0)  $ 
             & 0.23 $\pm$ 0.11 ${}^{+0.01}_{-0.04}$ 
                          & -0.09 $\pm$0.09
                          $\pm $0.01 
                          & 0.04 $\pm$ 0.07 \\
\hline
\end{tabular}
\caption{The experimental data of the direct CP asymmetry and the average. }
\end{center}
\end{table}
The CP asymmetries under the same assumption   
are  
\bea
A_{CP}^{0+} &\equiv & \frac{|A^{0-}|^2 - |A^{0+}|^2}{|A^{0-}|^2 + |A^{0+}|^2} 
             = - 2 r_A \sin\delta^A \sin\phi_3 \sim 0.0 ,\\
A_{CP}^{00} &\equiv & \frac{|\bar{A}^{00}|^2 - |A^{00}|^2}
                     {|\bar{A}^{00}|^2 + |A^{00}|^2} 
             = 2 r_C \sin\delta^C \sin\phi_3 ,\\
A_{CP}^{+-} &\equiv & \frac{|A^{-+}|^2 - |A^{+-}|^2}{|A^{-+}|^2 + |A^{+-}|^2} 
             = - 2 r_T \sin\delta^T \sin\phi_3 
               -   r_T^2 \sin2\delta^T \sin2\phi_3 ,\\
A_{CP}^{+0} &\equiv & \frac{|A^{-0}|^2 - |A^{+0}|^2}{|A^{-0}|^2 + |A^{+0}|^2} 
             = - 2 (r_T \sin\delta^T + r_C \sin\delta^C 
                  + r_A \sin\delta^A )\sin\phi_3 \nn \\
            & & ~~~~~~~~~~~~~~ 
               - 2 r_{EW} r_{T} \sin(\delta^T -\delta^{EW})\sin\phi_3
               - r_T^2 \sin2\delta^T \sin2\phi_3  \nn \\
            & & ~~~~~~~~~~~~~~
               + 2 r_{EW} r_{T} \sin\delta^T \cos\delta^{EW}\sin\phi_3 .
\eea
$A_{CP}^{0+}$ should be almost $0$ because of the smallness of $r_A$ and 
the data is also showing it. Up to the first order of $r$, there is a 
relation among the CP asymmetries as follows:
\bea
A_{CP}^{+0} = A_{CP}^{+-} - A_{CP}^{00} + A_{CP}^{0+}. 
\eea
The discrepancy of this relation is also caused by the cross term of $r_T$ 
and $r_{EW}$. If we can have more accurate data, this may also show an useful 
information about $r_{EW}$.   
Because $A_{CP}^{+-}$ is not so large, we can confirm that 
$r_{T}$ will not become so large value. 

In $B\rightarrow \pi^0 K_S $, we can also use some informations 
about time dependent CP asymmetry\cite{NQ}. 
The measurements for $B\rightarrow\pi^0 K_S$ are
\bea
\Gamma(B^0\rightarrow \pi^0 K_S) + \Gamma(\bar{B}^0\rightarrow \pi^0 K_S)
                             &\propto & (|A|^2 + |\bar{A}|^2 ), \\
\Gamma(B^0(t)\rightarrow \pi^0 K_S)
                        - \Gamma(\bar{B}^0(t)\rightarrow \pi^0 K_S)
         &\propto &
         (|A|^2 - |\bar{A}|^2 ) \cos\Delta m t  \nn \\
            & &  + 2 Im(e^{-2i\phi_1} A^* \bar{A}) \sin\Delta m t .
\eea
After neglecting the contribution from
annihilation diagram, we consider the following measurements under the same 
assumption, 
\bea
\frac{\tau^+}{\tau^0}
\frac{\Gamma(B^0\rightarrow \pi^0 K_S)
         + \Gamma(\bar{B}^0\rightarrow \pi^0 K_S)}
     {\Gamma(B^+\rightarrow \pi^+ K_S)} &\equiv& X \\ \label{X1}
                    &=& 1 - 2 r_{EW}\cos\delta^{EW} + r_{EW}^2 
                      + 2 r_C \cos\phi_3 \cos\delta^C , \nn 
                                     \\[5mm]
\frac{\tau^+}{\tau^0}
\frac{\Gamma(B^0(t)\rightarrow \pi^0 K_S)
        - \Gamma(\bar{B}^0(t)\rightarrow \pi^0 K_S)}
     {\Gamma(B^+\rightarrow \pi^+ K_S)} &\equiv &
             Y \cos\Delta m t - Z \sin\Delta m t \nn \\
          &=&
         (- 2 r_C \sin\phi_3 \sin{\delta^C} )\cos\Delta m t  \nn \\
            & &-\{ \sin2\phi_1 (1 - 2 r_{EW}\cos\delta^{EW} + r_{EW}^2)
\label{Y1}\\
    & &  + 2 r_C \sin{(\phi_3 + 2 \phi_1 )} \cos\delta^C \} \sin\Delta m t .
\nn
\eea
And we define them as follows:
\bea
X &=& 1 - 2 r_{EW} \cos\delta^{EW} + r_{EW}^2 
              + 2 r_C \cos\phi_3 \cos\delta^C, \\
Y &=& - 2 r_C \sin\phi_3 \sin{\delta^C}  \approx - A_{CP}^{00}, \\
Z &=& \sin{2\phi_1}(1 - 2 r_{EW}\cos\delta^{EW} 
                         + r_{EW}^2 )
            + 2 r_C \sin{(\phi_3 + 2 \phi_1 )} \cos\delta^C.
\eea
Eliminating $ r_C \cos\delta^C$ in $X$ and $Z$,
we find $\tan{\phi_3 }$ as a function of $r_{EW}$,
\bea
\tan{\phi_3 } = \frac{ \left\{ Z -  X \sin2\phi_1
                  \right\}}{\cos2\phi_1 \left\{ X - ( 1 
                             - 2 r_{EW}\cos\delta^{EW} + r_{EW}^2)
                    \right\}}.
\label{solve-phi3}
\eea
If one can estimate $r_{EW}$ at good accuracy, one can have a information
about $\phi_3$ by inputting the value of $\phi_1$ which is measured by 
$B\rightarrow J/\psi K_S$ at higher accurate experiment.   
It will be extracted from quantity of $O(0.01)$ because 
$r_C \sim 0.02 $ under the assumption in this work. Indeed, 
for the solution eq.(\ref{solution}), 
$2 r_C \cos\delta^C \cos\phi_3 \approx X 
- 1 + 2 r_{EW} \cos\delta^{EW} -r_{EW}^2 \sim -0.03$.  
So to use this measurements one may need some corrections from 
$K-\bar{K}$ mixing and the width difference. See Appendix. 
After extracting $\phi_3$ by the other modes, 
we can use to estimate or confirm 
how large is the electro-weak penguin contribution.   

In this paper, we discussed about  
a possibility of large electro-weak penguin contribution in 
$B\rightarrow K \pi$ from recent experimental data. 
The several relations among the branching ratios which realize when 
the contributions from tree type and electro-weak penguin are small 
compared with the gluon penguin do not satisfy the data. 
The difference comes from the $r^2$ 
terms and the main contribution comes from electro-weak penguin. 
We find that the contribution from electro-weak penguin 
may be larger than from tree diagrams to explain the experimental data. 
If the magnitude estimated from experiment is quite large compared 
with the theoretical estimation which is usually smaller 
than tree contributions\cite{NEU,FM,BBNS}, then it may be including 
some new physics effects. In this analysis, we find that who can have 
some contribution from new physics is 
the color favored electro-weak penguin type diagram which is the process 
$\pi^0$ goes out from $B-K$ current.

\section*{Acknowledgments}
We would like to thank Sechul Oh for many useful comments and discussions.    

\section*{Appendix: Some correction terms from the
$K-\bar{K}$ mixing and the width difference. }
In the discussion about the time dependent CP asymmetry of 
$B^0\rightarrow K_S \pi^0 $, we neglected the effect of CP violation 
in K meson system and
tiny width difference of $B_d$ mesons
because it is very small. The magnitude is $|\epsilon| = 2\times 10^{-3}$ and
$\Delta \Gamma_d /\Gamma_d$ is about $3 \times 10^{-3}$ which is estimated
in \cite{DHKY}.  If $r_C$ is the order, we must deal with the contributions.
In the Kon system, we define the $K_S$ and $K_L$ mesons as follows:
\bea
|K^0> &=& \frac{1 - \epsilon}{\sqrt{2}}
             \left[ |K_S> + |K_L> \right],  \\
|\bar{K}^0> &=& \frac{1 + \epsilon}{\sqrt{2}}
             \left[ |K_S> - |K_L> \right],
\eea
where $\epsilon$ is the parameter which shows the CP violation of K. Then
one needs several correction terms including $\epsilon$
and $\Delta \Gamma_d $ as the expansion parameters\cite{DHKY}
in $X$, $Y$ and $Z$
(Note that the definition of the sign of $\epsilon$ is different from it in
\cite{DHKY}. ) as follows:
\bea
X(t) &=& 1 - 2 r_{EW} \cos\delta^{EW} + r_{EW}^2
           + 2 r_C \cos\phi_3 \cos\delta^C
         + \cos2\phi_1 \sinh \frac{\Delta\Gamma_d t }{2} + \cdots, \\
Y &=& 2 r_C \sin\phi_3 \sin\delta^C - 2 Re(\epsilon) + O(\epsilon r_C),\\
Z &=& \sin2\phi_1 (1 - 2 r_{EW} \cos\delta^{EW} + r_{EW}^2 )
   + 2 r_C \sin(\phi_3 + 2 \phi_1) \cos\delta^C \nn \\
  & & ~~~  - 2 Im(\epsilon) \cos2\phi_1  + \sin2\phi_1 \cos2\phi_1
                \sinh \frac{\Delta\Gamma_d t }{2} + O(\epsilon r_C).
\eea
Here we neglected the terms of $\epsilon r $.
In addition to these terms, there are also a constant term and
a proportional term to $\sin^2(\Delta mt)$. $X$ has also a correction
term by $\epsilon$ but it is of order $\epsilon r $
so that one can neglect them.  However,
indeed, the correction terms has already been included in $\sin2\phi_1$
determined by $B\rightarrow J/\psi K_S$ \cite{DHKY}
and the value which subtracts from
$Z$ is effectively $ ( \sin2\phi_1 - 2 Im(\epsilon) \cos2\phi_1 + \cdots)$
so that one can neglect the effect in $Z$.

\end{document}